\documentclass[%
 reprint,
 superscriptaddress,
nopreprintnumbers,
nobibnotes,
 amsmath,amssymb,
 aps,
 prb,
]{revtex4-2}

\usepackage{graphicx}
\usepackage{dcolumn}
\usepackage{bm}

\usepackage{amsmath}
\usepackage{amssymb}
\usepackage[detect-all]{siunitx}
\usepackage{bm}
\usepackage{mathtools}
\usepackage{physics}

\usepackage[normalem]{ulem}
\usepackage{color}

\newcommand{\e}{\hat{\mathbf{e}}}
\newcommand{\m}{\hat{\mathbf{m}}}
\newcommand{\rr}{\mathbf{r}}
\newcommand{\q}{\mathbf{q}}

\begin{document}

\title{Photon correlation spectroscopy with heterodyne mixing based on soft-x-ray magnetic circular dichroism}

\author{Christopher Klose}
    \affiliation{Max Born Institute for Nonlinear Optics and Short Pulse Spectroscopy, Max-Born-Stra\ss e 2A, 12489 Berlin, Germany}
\author{Felix Büttner}%
    \affiliation{Helmholtz-Zentrum f\"ur Materialien und Energie GmbH, 14109 Berlin, Germany}%
    \affiliation{Department of Materials Science and Engineering, Massachusetts Institute of Technology, Cambridge, Massachusetts 02139, USA}
    \affiliation{National Synchrotron Light Source II, Brookhaven National Laboratory, Upton, NY, USA}%
\author{Wen Hu}
    \affiliation{National Synchrotron Light Source II, Brookhaven National Laboratory, Upton, NY, USA}%
\author{Claudio Mazzoli}
    \affiliation{National Synchrotron Light Source II, Brookhaven National Laboratory, Upton, NY, USA}%
\author{Geoffrey S. D. Beach}
    \affiliation{Department of Materials Science and Engineering, Massachusetts Institute of Technology, Cambridge, Massachusetts 02139, USA}
\author{Stefan Eisebitt}
    \affiliation{Max Born Institute for Nonlinear Optics and Short Pulse Spectroscopy, Max-Born-Stra\ss e 2A, 12489 Berlin, Germany}
    \affiliation{Technische Universit\"at Berlin, Institut f\"ur Optik und Atomare Physik, Stra\ss e des 17.\ Juni 135, 10623 Berlin, Germany}
\author{Bastian Pfau}%
 \email{bastian.pfau@mbi-berlin.de}
\affiliation{Max Born Institute for Nonlinear Optics and Short Pulse Spectroscopy, Max-Born-Stra\ss e 2A, 12489 Berlin, Germany}%

\date{\today}

\begin{abstract}
Many magnetic equilibrium states and phase transitions are characterized by fluctuations. Such magnetic fluctuation can in principle be detected with scattering-based x-ray photon correlation spectroscopy (XPCS). However, in the established approach of XPCS, the magnetic scattering signal is quadratic in the magnetic scattering cross-section, which results not only in often prohibitively small signals but also in a fundamental inability to detect negative correlations (anticorrelations). Here, we propose to exploit the possibility of heterodyne mixing of the magnetic signal with static charge scattering to reconstruct the first-order (linear) magnetic correlation function. We show that the first-order magnetic scattering signal reconstructed from heterodyne scattering now directly represents the underlying magnetization texture. Moreover, we suggest a practical implementation based on an absorption mask rigidly connected to the sample, which not only produces a static charge scattering signal but also eliminates the problem of drift-induced artificial decay of the correlation functions. Our method thereby significantly broadens the range of scientific questions accessible by magnetic x-ray photon correlation spectroscopy. 
\end{abstract}

\maketitle

\section{Introduction}
Thermal and quantum fluctuations are of fundamental importance for the physics and function of many magnetic materials. Fluctuations particularly emerge during phase transitions as, for instance, critical fluctuations close to the magnetic ordering temperature or Barkhausen noise during a ferromagnetic reversal---both representing standard models of magnetism. More recent examples comprise the fluctuation dynamics of frustrated systems such as artificial spin-ice materials \cite{kapaklis_NN_2014,perrin_N_2016,farhan_SA_2019,chen_PRL_2019}, spin and charge density fluctuations in high-T\textsubscript{c} superconductor materials \cite{ghiringhelli_S_2012,mitrano_SA_2019,arpaia_S_2019} and a fluctuation-mediated topological phase transition \cite{buttner_NM_2020}. 
Technologically, fluctuations play a decisive role for the stability of magnetic recording media where thermal activation leads to the switching of the magnetic area representing a logic bit \cite{pfau_JoAP_2017}. The resulting minimum size of the magnetic grains for long-term stability gives rise to the superparamagnetic limit for the recording density \cite{bean_JAP_1959, richter_JPDAP_2007}.

Correlation-based x-ray scattering methods (typically referred to as x-ray photon correlation spectroscopy, XPCS) provide a very general and direct way to study fluctuations in the time domain \cite{sutton_CRP_2008, madsen_SLSaFLAPIaSA_2018}. Generally, these coherent x-ray methods rely on the analysis of temporal speckle correlations in the scattered intensity.
In contrast to its visible-light counter part, called dynamical light scattering (DLS), XPCS provides access to optically opaque materials and gains nanometer-scale and even atomic-scale sensitivity from the small x-ray wavelength \cite{ruta_PRL_2012, leitner_NM_2009}.

So far, applications of XPCS are still rare in magnetism research. Examples---also including methodically related investigations based on resonant elastic x-ray scattering---comprise the dynamics and memory effects of magnetic domains \cite{konings_PRL_2011, pierce_PRL_2003,chesnel_NC_2016} and charge and spin density waves \cite{shpyrko_N_2007,chen_NC_2019} as well as artificial spin-ice dynamics \cite{chen_PRL_2019} and orbital-domain dynamics \cite{kukreja_PRL_2018,turner_NJP_2008}. These experiments were carried out on the timescale of seconds or even quasi-statically. For magnetic samples, the x-ray magnetic circular dichroism (XMCD) in the soft-x-ray range is typically exploited as contrast. The time limitation arises, on the one hand, from the low magnetic scattering signal compared to the scattering from electron-density variations and, on the other hand, from the slow area detectors available so far, particularly for soft x rays. While experiments at x-ray free-electron lasers have demonstrated magnetic XPCS on the ultrafast timescale (ps to ns) \cite{seaberg_PRL_2017,esposito_APL_2020, seaberg_PRR_2021}, a sizable gap in the application time window remains. 

This situation is supposed to improve in the near future as the soft-x-ray detector technology is about to be significantly advanced \cite{bergamaschi_SRN_2018, marras_JSR_2021, desjardins_ACP_2019} and forth-generation synchrotron-radiation sources are coming in operation, providing almost fully coherent soft x rays of much higher intensity. However, the progression towards much shorter timescales will still be hampered by the low magnetic scattering cross-section. The scattering geometry is typically designed in a way that the weak, magnetically scattered intensity is well separated from charge-based contributions in the scattering space, which allows for an almost background-free detection of the magnetic signal \cite{pfau_NC_2012}. Nevertheless, the intensity of this homodyne magnetic scattering is typically $10^2$ to $10^4$ times smaller than the charge-based scattering (corresponding to a $10$ to  $10^2$ times smaller magnetic absorption cross-section). Ultimately, this signal will become too small for short-term detection, in particular, as faster dynamics are often additionally accompanied by a further reduction of the magnetic signals due to a general loss of magnetization as well as due to smaller relevant length scales and correspondingly higher scattering angles \cite{buttner_NM_2020}. 

Here, we propose to extend the application frame of magnetic XPCS by a change of the scattering geometry to a heterodyne detection scheme. Heterodyne XPCS is achieved by mixing the scattered signal wave from the fluctuating sample with a temporally constant reference wave \cite{gutt_PRL_2003}. 
As one of the main differences to the homodyne detection, the interference with the reference wave preserves the scattering amplitude of the signal wave and its relative phase to the reference \cite{cummins__1977}. Heterodyning is a standard technique in DLS \cite{cummins__1977,schatzelk_APB_1987} and is also exploited by coherent x-ray methods. In particular, heterodyne mixing is applied in XPCS velocity measurements of slow-moving materials and particles \cite{livet_JAC_2007, ehrburger-dolle_M_2012, ulbrandt_NP_2016} as well as in holographic x-ray imaging \cite{eisebitt_N_2004,pfau_SLSaFL_2015}. When exploiting the XMCD, a coherent mixing between magnetic scattering and charge scattering naturally occurs for x rays with circular polarization \cite{eisebitt_PRB_2003, kortright_JESRP_2013, Wang_PRL_2012}. If the charge scattering signal is static, it can serve as reference to realize a heterodyne XPCS experiment. If the charge-based reference signal is stronger than the magnetic signal, the latter will be coherently amplified. As a result, heterodyne detection of the magnetic scattering is supposed to increase the XPCS signal to overcome instrumental noise, in a similar way as already demonstrated for reference-assisted coherent diffractive imaging \cite{kim_OEO_2014,huang_AA_2020}.

Even as important in the context of magnetic systems, heterodyne scattering also provides access to first-order correlations and therefore allows to detect anticorrelations. In practice, this means that an inversion of the magnetic structure such as a magnetic switching event appears in the heterodyne XPCS signal while remains hidden in the homodyne case.

In soft-x-ray experiments, it is straightforward to obtain the static reference signal from an aperture which is fabricated directly on the sample or mounted in its close proximity \cite{eisebitt_PRB_2003}. Such an aperture can be well adapted in size and shape to the experiment and provides excellent temporal stability \cite{chen_NC_2019}.

We will derive the first-order, two-time correlation function for this particular experimental geometry. Our results can be transferred easily to similar setups. Our goal is to prove that the temporal first-order speckle correlations directly reflect correlations of the changing real-space magnetization texture. We will show under which conditions and how well this direct relation is achieved. The analytic derivation is complemented by Monte-Carlo-like simulations to provide an error estimate of the method and demonstrate the sensitivity gained from the stability of the mask-based approach.

\section{General considerations}
X rays provide sensitivity to the magnetization of a material via a dichroism, i.e., via a dependence of material's x-ray absorption on the x-ray polarization state. In more general terms, the effect is described by a polarization-dependent atomic x-ray scattering factor $f^n(\omega)$ of the magnetized atom $n$. As the dichroism is particularly pronounced at electronic resonances, the magnetization contrast is typically obtained in an element specific way. In the absence of any charge (natural) dichroism, the magnetic dichroism is described via the expansion of the elastic resonant atomic scattering factor into charge and magnetic scattering amplitudes \cite{hannon_PRL_1988,stohr__2006,kortright_JESRP_2013}:
\begin{eqnarray}
f^n(\omega) =\; && (\e^*_2 \cdot \e_1) f^n_\mathrm{c}(\omega) - \mathrm{i}\,(\e^*_2 \times \e_1) \cdot \m^n \, f^n_\mathrm{m1}(\omega) \nonumber\\
&& +\, (\e^*_2 \cdot \m^n)(\e_1 \cdot \m^n) \, f^n_\mathrm{m2}(\omega) + \ldots
\end{eqnarray}
Here, the charge ($f^n_\mathrm{c}(\omega)$) and magnetic ($f^n_\mathrm{m1}(\omega)$, $f^n_\mathrm{m2}(\omega)$) atomic scattering factors contain the sums of the transition probabilities of the atomic excitation and decay processes involved in the resonant scattering process. The polarization dependence is given by specific vector products of the complex unit polarization vectors of the incident ($\e_1$) and the scattered ($\e_2$) x rays \cite{detlefs_EPJST_2012}, as well as $\m^n$, the unit vector along the magnetic polarization axis of the scattering atom. The asterisk symbol ($\ast$) indicates the complex conjugation. The first magnetic scattering term describes the XMCD and the second term the linear dichroism. 

Focusing on ferromagnetic materials (including ferrimagnets with ferromagnetic sublattices or sublayers), the XMCD typically dominates the magnetic response if the magnetization is suitably aligned to the x-ray beam. In addition, spatial spin textures in these materials appear on much larger length scales than atomic distances and, thus, only the forward scattering amplitude of the atomic cross-section has to be considered. Using circularly polarized x rays, the scattered intensity with a scattering vector $\q$ is then provided by the superposition (i.e.\ the summation) of the scattered waves of all atoms at positions $\rr_n$ in the coherence volume in the sample, i.e., the magnitude squared of the structure function:
\begin{eqnarray}
 I_\pm(\q)\ &\propto& \left| \sum_n f^n e^{\mathrm{i} \q \rr_n} \right|^2 \\
  &\propto& \left| \sum_n (f^n_\mathrm{c} \pm \e_k \cdot \m^n  f^n_\mathrm{m1}) e^{\mathrm{i} \q \rr_n} \right|^2. \label{eq:scatt intensity full}
\end{eqnarray}
The XMCD polarization dependence reduces to $\pm \e_k \cdot \m^n$ where $\e_k = \mathbf{k}/\left|\mathbf{k}\right|$ denotes the x-ray propagation direction and the different signs distinguish left or right circularly polarized x rays. The contrast is, thus, maximized for a parallel alignment between the x-ray beam and the magnetization.

The sum in Eq.~\eqref{eq:scatt intensity full} can be readily divided into charge and magnetic structure functions:
\begin{eqnarray}
 I_\pm(\q) \propto \left| F_\mathrm{c} \pm F_\mathrm{m} \right|^2, 
\end{eqnarray}
where $F_\mathrm{c} = \sum_n f^n_\mathrm{c} \exp(\mathrm{i} \q \rr_n)$ and $F_\mathrm{m} = \sum_n  \e_k \cdot \m^n  f^n_\mathrm{m1} \exp(\mathrm{i} \q \rr_n)$. After expansion, this gives:
\begin{eqnarray}
I_\pm(\q)\; && \propto F^\ast_\mathrm{c} F_\mathrm{c} \pm 2\Re [ F^\ast_\mathrm{c} F_\mathrm{m}] + F^\ast_\mathrm{m}F_\mathrm{m}. \label{eq:circular_scatt}
\end{eqnarray}
The scattered x-ray intensity is composed of a purely charge-based and a purely magnetic homodyne contribution as well as the heterodyne term as interference between charge and magnetic scattering. The mixed term vanishes for linear polarization which can be perceived as a superposition between left and right circularly polarized photons \cite{eisebitt_PRB_2003, kortright_JESRP_2013}:
\begin{eqnarray}
I_\mathrm{lin}(\q) \propto F^\ast_\mathrm{c} F_\mathrm{c} + F^\ast_\mathrm{m}F_\mathrm{m}.
\end{eqnarray}
In this case, the scattered intensity is composed of the incoherent sum of charge and magnetic scattering. In such an experiment, the magnetic scattering can mostly be isolated by different approaches:
\begin{itemize}
    \item Subtracting a separately recorded pure charge signal if available.
    \item Reducing the charge scattering signal as much as possible by using uniform, high-quality magnetic films and substrates and a very smooth, e.g., Gaussian illumination function.
    \item Tailoring the size and shape of the topographic sample features (such as lithographic structures or apertures) in a way that charge and magnetic scattering are well separated in Fourier space. In addition, the charge scattering can be blanked out by suitably designed beamstops or favorable positioning of the detector \cite{weder_SD_2020}.
\end{itemize}
The isolated magnetic scattering can then be readily analyzed by standard XPCS methods yielding second-order (intensity) correlation functions \cite{sutton_CRP_2008, madsen_SLSaFLAPIaSA_2018}.

In contrast, the additional heterodyne term in Eq.~\eqref{eq:circular_scatt} also provides access to the first-order correlation function. The pure charge term can be eliminated by a separately recorded charge scattering signal as described above. The pure magnetic term now appears to be small compared to the mixed term and can be neglected.  

In the following, we will derive Eq.~\eqref{eq:circular_scatt} for the specific case of a uniform magnetic film covered by an x-ray mask which is located in close proximity to the sample. The mask contains a single aperture and is opaque to x rays elsewhere. This topographic structure, the magnetic textures in the film considered and the soft-x-ray wavelength are large compared to atomic distances. We, therefore, derive the equations in the wave picture using the continuous (dichroic) refractive index to describe the sample. From the equation of the scattered intensity, we will then derive the first-order, two-time correlation function, describing the dynamics of the magnetic system. 
\section{Mask-based heterodyne setup}
\label{SI:sec:fth}

\begin{figure}[tb]
    \centering
    \includegraphics[width=0.9\linewidth]{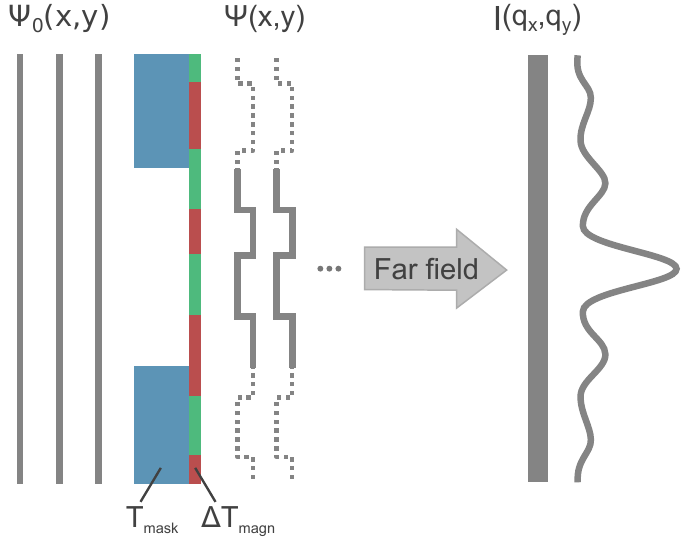}
    \caption{\textbf{Geometry of the proposed transmission experiment for heterodyne detection of the magnetic scattering signal.} Symbols are explained in the text.
    }
    \label{Fig:Geometry}
\end{figure}

The geometry of the generic coherent scattering experiment in transmission is shown in Fig.~\ref{Fig:Geometry}. 
The sample is illuminated by a coherent plane-wave x-ray beam $\Psi_0(\rr)=\left|\Psi_0\right|\exp(\mathrm{i}(kz-\omega t))$ propagating in $z$ direction with frequency $\omega$ and wavenumber $k = 2\pi/\lambda$. The scattering contrast arises from a spatially varying refractive index $n(\omega,\rr)$ of the sample, where $\rr$ denotes the position in the sample plane. For ferromagnetic samples, the dichroic complex refractive index $n(\omega,\rr,z,\sigma)$ for right ($\sigma=1$) and left ($\sigma=-1$) circularly polarized light is commonly written as \cite{pfau_SLSaFL_2015}:
\begin{eqnarray}
    n(\omega,\rr,z,\sigma)= 1-\delta_0\left(\omega,\rr,z\right)+\mathrm{i}\beta_0\left(\omega,\rr,z\right)\nonumber \\
    + \left(-\Delta\delta\left(\omega,\rr,z\right) \right.
    + \left. \mathrm{i}\Delta\beta\left(\omega,\rr,z\right)\right)\sigma\e_k\cdot\m(\rr,z), 
    \label{eq:refractive index}
\end{eqnarray}
where $\delta_0$ and $\beta_0$ denote the optical constants for unpolarized or linearly polarized light. The additional constants $\Delta\delta$ and $\Delta \beta$ refer to the XMCD. The optical constants are directly derived from the atomic scattering factors and the material's atomic density \cite{attwood__1999}. The strength of the XMCD depends on the local magnetization direction $\m(\rr,z)$ with respect to the orientation of the light's propagation direction $\e_k = \e_z$. Using the projection approximation \cite{nugent_AP_2010,pfau_SLSaFL_2015}, the exit wave $\Psi_\mathrm{exit}(\rr)$ is a product of $\Psi_0(\rr)$ and the sample transmittance $T(\rr)$:
\begin{eqnarray}
    \label{eq:Phi_exit_r}
    \Psi_\mathrm{exit}(\rr) &=& T(\rr)\Psi_0(\rr)\\
    \label{eq:transmission}
    T(\rr) &=& \exp\left[\mathrm{i}k\int_{-d}^0 (n(\rr,z)-1)\,\mathrm{d}z \right],
\end{eqnarray}
where $n(\rr,z)-1$ is integrated over the sample thickness $d$. This equation is valid for magnetic and non-magnetic samples. The exit wave at $z=0$ propagates in free space to the detector located in the Fraunhofer far-field region. The scattered wave front $\Psi_\mathrm{det}(x_\mathrm{det},y_\mathrm{det})$ at the detector is therefore represented by the Fourier transform of $\Psi_\mathrm{exit}$ (trivial phase terms are omitted):
\begin{eqnarray}
    \label{eq:Phi_exit_q}
    \Psi_\mathrm{det}(x_\mathrm{det},y_\mathrm{det}) = \mathcal{F}\left[\Psi_\mathrm{exit}\right](\q) .
\end{eqnarray}
Here, $\q=\left(q_x,q_y\right)\approx (k/z_0) \left(x_\mathrm{det},y_\mathrm{det}\right)$ represents a reciprocal-space coordinate, which is linked to the real-space coordinates at the detector at distance $z_0$ from the sample. The initial factor $\left|\Psi_0\right|$ and any spatially uniform contributions of $T$ only influence the integrated intensity detected without generating contrast. We, hence, focus our discussion on the spatially varying part of $T(\rr)$.

Our generic sample comprises two distinct layers, the mask and the actual magnetic film, with transmission functions $T_\mathrm{mask}(\rr)$ and $T_\mathrm{s}(\rr)$, respectively. Assuming thin layers compared to the lateral size of the aperture, we can again use the projection approximation: $T(\rr)=T_\mathrm{mask}(\rr)T_\mathrm{s}(\rr)$. Additionally, we make the following assumptions:
\begin{enumerate}
    \item  The mask is a circular hole in a fully opaque layer.
    In particular, $T_\mathrm{mask} (\rr)= T_\mathrm{mask}(-\rr)$ and $\mathcal{F}[T_\mathrm{mask}]$ is real.
    
    \item The sample is sufficiently thin such that the exponential function in Eq.~\eqref{eq:transmission} can be approximated by the Taylor expansion up to linear order:  $T_\mathrm{s}(\rr)=1+\Delta T_\mathrm{s}(\rr)$.
    
    \item For homogeneous films, the contrast $\Delta T_\mathrm{s}$ is generated by a single, scalar, real-valued quantity $C(\rr)$. In case of magnetic samples, this quantity is the product of helicity, light propagation direction, and local magnetization unit vector: $C(\rr) = (\sigma/d) \int  \e_k \cdot \m(\rr,z) \,\mathrm{d}z$. The integration is carried out over all magnetized layers which are resonantly probed. The contrast is entirely magnetic and given by the $z$ component of the magnetization ($ m_z(\rr)$). The sample transmission is then given by the product of $C$ and the (complex) contrast factor $z_C$ corresponding to the quantity $\sigma \Delta T_\mathrm{magn} \equiv \Delta T_\mathrm{s} = z_C C(\rr) = \sigma z_C m_z(\rr)$. For XMCD contrast, $z_C=\mathrm{i}kd (-\Delta\delta + \mathrm{i}\Delta\beta )$. Note that any topography of the magnetic film can be represented by adapting $T_\mathrm{mask}$.
\end{enumerate}
The detector records the signal intensity ($I(\q,\sigma))$, i.\,e., the squared absolute value of $\Psi_\mathrm{det}$:
\begin{align}
    \label{eq:I_q_s}
    I(\q,\sigma) \propto 
    \left|\mathcal{F}\left[ T_\mathrm{mask}(\rr)\left(1+\sigma\Delta T_\mathrm{magn}(\rr)\right)\right]
    \right|^2 .
\end{align}
For an XPCS experiment, this intensity is sampled in time frames in order to monitor the dynamic behavior of the sample. In the next step, we will derive the general two-time correlation for two intensity measurements $I_1$ and $I_2$ and times $t_1$ and $t_2$, respectively.

\section{Derivation of first-order two-time-correlation function}
\label{SI:sec:correlation_function}

As discussed earlier, heterodyne magnetic scattering provides an interference term which is linear in the sample's magnetization. Therefore, it is our explicit goal to derive an equation for the two-time correlation $c_\textrm{q}\left(t_1,t_2\right)$ of the scattered intensity which yields the same result as a correlation of the underlying magnetic configurations in real-space would do. This correlation function will go beyond the information provided by classical (homodyne) XPCS experiments as it will also resolve anticorrelations and, e.g., provide information on magnetic switching events which for many systems are characteristic for the underlying magnetization dynamics.

We first recall the equation of the real-space correlation function:
\begin{align}
c_\textrm{r}(t_1,t_2)=\frac{\langle m_1,m_2 \rangle}{\lVert m_1\rVert \lVert m_2\rVert},
\end{align}
where $m_i=m_z(\rr,t_i)\propto T_\text{mask}\Delta T_\text{magn}$ is the out-of-plane magnetization in the field of view defined by the mask transmission function $T_\text{mask}$. Numerically, the brackets $\langle.,.\rangle$ indicate the point-by-point scalar product and $\lVert.\rVert$ the corresponding norm. 

According to Plancherel's theorem, scalar products are identical in Fourier space and in real space. Therefore,
\begin{align}
\frac{\langle m_1,m_2 \rangle}{\lVert m_1\rVert \lVert m_2\rVert} = \frac{\langle \mathcal{F}[m_1],\mathcal{F}[m_2] \rangle}{\lVert \mathcal{F}[m_1]\rVert \lVert \mathcal{F}[m_2]\rVert}.
\end{align}
Normally, $\mathcal{F}[m_z]$ is inaccessible from scattering data since the complex phase of the Fourier transform is lost in the detection process. However, the heterodyne mixing of topographic and magnetic signals allows to partially recover the sign of the wavefront from the interference of the mask scattering and the magnetic scattering.

We start with recording scattering patterns of opposite helicity ($\sigma$) and calculate the sum ($I_\mathrm{sum}(\q) = I(\q,+1) + I(\q,-1)$) and difference ($I_\mathrm{diff}(\q) = I(\q,+1) - I(\q,-1)$) of theses patterns based on Eq.~\eqref{eq:I_q_s}:
\begin{eqnarray}
    \label{eq:I_diff}
    \nonumber
   I_\mathrm{diff}(\q) &\propto& \mathcal{F}\left[\left(T_\mathrm{mask}\Delta T_\mathrm{magn}\right)(\rr)\right]
     \cdot \mathcal{F}^*\left[T_\mathrm{mask}(\rr)\right]\\
     &+& \mathcal{F}^*\left[\left(T_\mathrm{mask}\Delta T_\mathrm{magn}\right)(\rr)\right]
     \cdot \mathcal{F}\left[T_\mathrm{mask}(\rr)\right]\\
     \label{eq:I_sum}
     \nonumber
    I_\mathrm{sum}(\q)  &\propto& \left|\mathcal{F}\left[T_\mathrm{mask}(\rr)\right]\right|^2\\
    &+& \left|\mathcal{F}\left[\left(T_\mathrm{mask}\Delta T_\mathrm{magn}\right)(\rr)\right]\right|^2 .
\end{eqnarray}
The difference contains the magnetic--charge mixed terms of interest.
For a symmetric object aperture, $\mathcal{F}\left[T_\mathrm{mask}(\rr)\right]$ is a real-valued function. Moreover, the mask scattering $\left|\mathcal{F}\left[T_\mathrm{mask}(\rr)\right]\right|^2$ is much stronger than the other terms contributing to $I_\mathrm{sum}(\q)$. We can therefore simplify the expressions to:
\begin{eqnarray}
    I_\mathrm{diff}(\q) &\propto& \Re \left( z_C \mathcal{F}\left[m_z(\rr)\right]\right)\mathcal{F}\left[T_\mathrm{mask}(\rr)\right] \label{eq:I_diff_2}\\
    I_\mathrm{sum}(\q)  &\propto& \mathcal{F}\left[T_\mathrm{mask}(\rr)\right]^2.
\end{eqnarray}
Hence, if we define $S=I_\mathrm{diff}(\q)/\sqrt{I_\mathrm{sum}(\q)}$, we can calculate the correlation via
\begin{eqnarray}
c_\mathrm{q}(t_1,t_2)&=&\frac{\langle S_1,S_2 \rangle}{\lVert S_1\rVert \lVert S_2\rVert} \label{Eq:c_q_1}\\
&=& \frac{ \left\langle\Re\left( z_C \mathcal{F}\left[m_1\right]\right),
\Re\left( z_C \mathcal{F}\left[m_2\right]\right)\right\rangle
}{
\lVert \Re\left( z_C \mathcal{F}\left[m_1\right]\right)\rVert
\lVert \Re\left( z_C \mathcal{F}\left[m_2\right]\right)\rVert}.
\label{SI:eq:c_q_2}
\end{eqnarray}
This expression is identical to $c_\mathrm{r}(t_1,t_2)$ if we assume that the correlation of the real part of $\mathcal{F}[m_z]$ is approximately equivalent to the complex correlation. In principle, only when $z_C$ has non-zero imaginary part, information of the imaginary part of $\mathcal{F}\left[m_z\right]$ is transferred to the real part. We further investigate and discuss this approximation in  Sec.~\ref{SI:sec:real_part_approximation}.

As the sum ($I_\mathrm{sum}$) only depends on the static topography function of the aperture, it can be determined once at the beginning of the experiment. The difference can then be calculated from one helicity measurement alone as $I_\mathrm{diff}(\q) = 2\sigma I(\q,\sigma) - \sigma I_\mathrm{sum}(\q)$. A helicity switch within the time series of an XPCS recording is not necessary. The experiment greatly benefits from avoiding the helicity switching during a time series for a series of reasons. First, because it naturally saves a significant amount of time, second, because the time series can be kept continuous and, thus, can be recut or rebinned as needed, and, last but not least, because it dramatically improves the measurement stability. However, the condition remains, that $I_\mathrm{sum}$ remains a correct approximation of the topography over time.

\section{Comparison of the Fourier-space and real-space correlation}
\label{SI:sec:real_part_approximation}

\begin{figure}[htb]
    \centering
    \includegraphics[width=\linewidth]{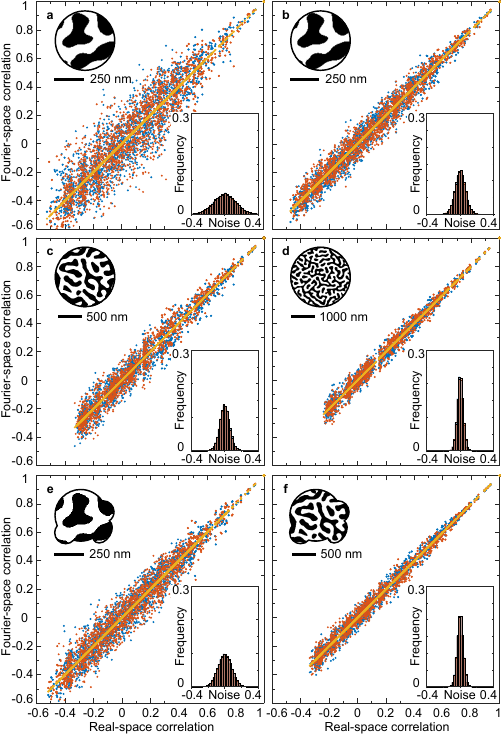}
    \caption{\textbf{Noise estimation of the Fourier-space correlation.} All panels show the connection between the real-space and Fourier-space two-time correlation based on a set of binary, maze-like domain patterns. Deviations between both correlations are considered as noise. Magnetic contrast $z_C$ is simulated as equal absorption and phase contrast $z_C = 1+ \mathrm{i}$ (yellow), pure absorption $z_C = 1$ (blue), or pure phase $z_C = \mathrm{i}$ (red) contrast in the transmission function, respectively. An exception is (b) where instead of pure absorption and phase contrast, $z_C = 1 + 0.6\,\mathrm{i}$ and $0.6+\mathrm{i}$, respectively. (a)--(d), Results when using a circular aperture mask with a diameter of (a,b) \SI{490}{nm}, (c) \SI{1225}{nm} and (d) \SI{2450}{nm}. (e),(f), Results for an asymmetric mask. The mask areas in (a) and (e) as well as (c) and (f) are equal. The insets of each panel show an example configuration of the magnetic pattern and the noise distribution of the data in the main plot (excluding the data for $z_C = 1+ \mathrm{i}$). While the histogram insets are based on the entire simulation data, only a reduced number of data points is shown in the scatter plots for illustration purposes.
    }
    \label{SI:fig:Simulation}
\end{figure}

According to Eq.~\eqref{SI:eq:c_q_2}, the heterodyne intensity correlation function is only sensitive to the real part of the Fourier transform of the magnetic transmission function. The imaginary part is lost in this procedure. Consequentially, the question arises how large the deviation between the Fourier-space and real-space correlation caused by this loss of information actually is. We aim to answer this question by comparing both correlation functions using examples of typical magnetic domain states. 

The simulation is based on a Monte-Carlo-like approach where magnetic domain patterns are randomly cropped from a $\SI{10}{\micro m}\cross \SI{10}{\micro m}$ magnetic force microscope image of a Co/Pt multilayer with out-of-plane anisotropy. We use a binary representation of the domains with opposite magnetization showing maze-like shape and an in-plane correlation length (twice the domain width) of $\lambda_\mathrm{D}\simeq\ \SI{190}{nm}$. The pixel size corresponds to \SI{4.9}{nm}. The cropping function is a binary circular aperture mask. We vary its diameter from \SI{490}{nm} to \SI{2450}{nm}. In addition, we use asymmetric aperture masks with mask areas equal to the spherical ones. In addition to the size and shape of the aperture, we vary the complex contrast factor as $z_C = 1,\, \mathrm{i}, \, \text{and}\, 1+\mathrm{i}$. For the smallest aperture, we also use $z_C = 1+ 0.6 \mathrm{i},\, \text{and}\,  0.6 + \mathrm{i}$. The real-space correlations were calculated directly from the magnetization patterns. The heterodyne Fourier-space correlation was derived using Eq.~\eqref{Eq:c_q_1}.

In Fig.~\ref{SI:fig:Simulation}, we present the results of our simulations with each panel combining the results for a selected aperture size and shape and different values of $z_C$. Each data point corresponds to the correlation of two randomly selected domain configurations.

Specifically, the plots compare the correlation results obtained in real space (abscissa) and Fourier space (ordinate). In the ideal case, the Fourier-space correlations should match the real-space target values and all points should be distributed along the diagonal of the plots. The deviation from the diagonal can either be conceived as an \emph{error} of the heterodyne correlation when comparing two specific states of the magnetic system or as additional \emph{noise} when averaging the correlation over many states. The corresponding noise distributions are shown as inset in each plot.
 
We make four important observations in our simulation. First, as expected, the correlation based on the heterodyne scattering correctly reproduces negative correlation, i.e., anticorrelations. 

The second observation concerns the dependence on the contrast factor $z_C$. When absorption and phase contrast are equally strong ($z_C = 1+i$), the Fourier-space correlation is accurate. For an increasing imbalance of absorption and phase contrast (Fig.~\ref{SI:fig:Simulation}b) up to even pure absorption ($z_C=1$) or phase contrast ($z_C=i$) (Fig.~\ref{SI:fig:Simulation}a) the mean deviation of the Fourier-space correlation also increases. The width of the noise distribution is independent on whether the real or imaginary part of $z_C$ dominates.
This observation can be understood from Eq.~\eqref{SI:eq:c_q_2}. The Fourier transform of any real-valued function $m_z(\rr)$ contains the inversion symmetric part of $m_z(\rr)$ in the real part and the antisymmetric part of $m_z(\rr)$ in the imaginary part. The complex multiplication with $z_C$ decides which information is transferred into the real part, which is conserved in the correlation function. If $z_C$ is only real (imaginary) only the real (imaginary) part of $\mathcal{F}\left[m_z(\rr)\right]$ is conserved. Conversely, if $z_C$ has equal real and imaginary parts, the full information on $m_z(\rr)$ is taken into the correlation function.
Fortunately, changing the contrast to mixed absorption and phase contrast can be achieved experimentally by fine-tuning the energy of the incident x-ray beam \cite{scherz_PRB_2007}. 

The third observation is related to the size of the aperture defining the x-ray illumination on the sample. When the size of the area probed becomes larger, it becomes less likely that a pattern exhibits a particular symmetry. This is illustrated in Figs.~\ref{SI:fig:Simulation}c,d. As more and more domains are captured by the aperture, changes between two states tend to be represented more and more equally in the real and imaginary parts of the Fourier transform of these patterns. As a result, the Fourier space correlation function converges to the real-space correlation result in the limit of large probe areas. Similar effects were already discussed in DLS experiments on non-ergodic media \cite{joosten_PRA_1990,pusey_PASMaiA_1989} and small scattering volumes \cite{cummins__1977,voigt_PASMaiA_1994}, and recently also theoretically in XPCS experiments \cite{bikondoa_C_2020} where a strong dependence of the correlation functions on the beam size and illuminated area was observed.

At last, we observe that also the aperture's shape influences the noise of the Fourier-space correlation.
Comparing apertures with equal area, apertures with asymmetric shapes show lower noise (given as the root mean square, RMS, of the noise distribution) compared to circular, symmetric apertures (Fig.~\ref{SI:fig:Simulation}a vs.\ e and Fig.~\ref{SI:fig:Simulation}c vs.\ f). We can understand this observation by going back to Eq.~\eqref{eq:I_diff_2}, where we applied the assumption of a symmetric aperture, allowing us to separate magnetic and charge scattering in the mixed term. When using an asymmetric mask this separation is not possible anymore as $\mathcal{F}\left[T_\mathrm{mask}(\rr)\right]$ is not a purely real-valued function anymore. However, the mixing (i.e., the multiplication term $\mathcal{F}\left[T_\mathrm{mask}(\rr)\right]\mathcal{F}\left[m_z(\rr)\right]$) now transfers the information of the imaginary part of $\mathcal{F}\left[m_z(\rr)\right]$ to the real part of the product, similar to the effect of a complex $z_C$ described above. This gain improves the Fourier-space correlation for asymmetric masks compared to symmetric ones.

\begin{figure}[htb]
    \centering
    \includegraphics[width=\linewidth]{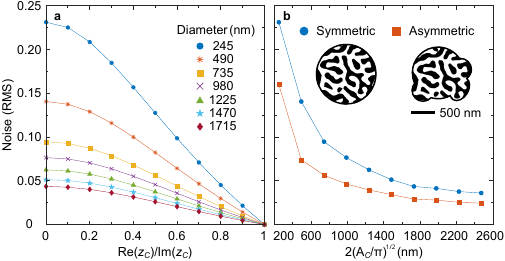}
    \caption{\textbf{Influence of magnetic contrast and aperture size on the noise distribution.} Root mean square (RMS) of the noise distribution derived from the deviation between the Fourier-space and real-space correlation for different experimental configurations. Sets of magnetic patterns were analyzed as (a) a function of the diameter of the circular aperture and the ratio between absorption and phase contrast, and (b) a function of the aperture size $2 \sqrt{A_C/\pi}$ ($A_C$ denotes aperture area) for a symmetric and an asymmetric aperture shape. The magnetic configurations in the insets exemplify the different aperture shapes.} 
    \label{fig:size}
\end{figure}

We summarize the simulation results on the noise of the heterodyne correlation in Fig.~\ref{fig:size}. We again use the RMS of the noise distribution as indicator. In Fig.~\ref{fig:size}a, we show how the noise decreases when the ratio between real and imaginary part of the contrast factor ($\Re (z_C) / \Im (z_C) $) increases from zero to unity. The influence of the aperture size as already described above is also evident. In Fig.~\ref{fig:size}b, we show the noise of the heterodyne correlation for different aperture shapes and sizes given as the square root of the aperture's area $A_C$. While we observe significant noise for small aperture sizes, the noise RMS quickly decreases with increasing aperture size and drops below \SI{5}{\percent} at an aperture size of approximately \num{8} times the in-plane correlations length ($\lambda_\mathrm{D}$) of the domains. In the case of asymmetric apertures the noise RMS is systematically lower compared to circular apertures and falls below \SI{5}{\percent} already at \num{\sim 4.7}$\lambda_\mathrm{D}$.
Experimentally, the probed area is often even larger than these sizes and we conclude, that practically the heterodyne first-order correlation function almost resembles the real-space correlations. We note, however, that this result was obtained from simulations based on isotropically ordered magnetic textures. In the case of, e.g., magnetic nanopatterns, additional, sample-specific simulations are needed to verify this relation.

\section{Advantages of a drift-free setup}
\label{SI:sec:Drift_approximation}

\begin{figure}[htb]
    \centering
    \includegraphics[width=0.8\linewidth]{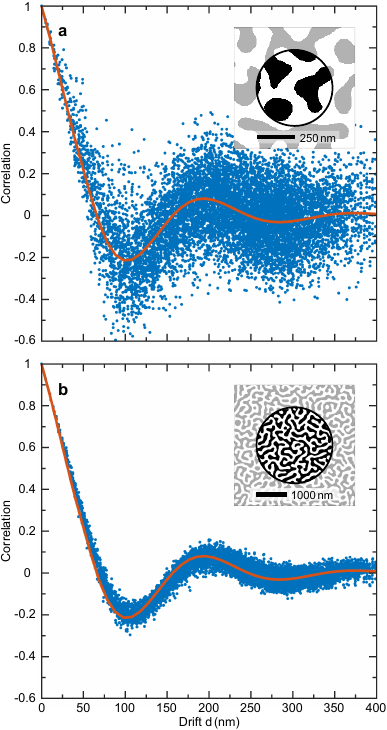}
    \caption{\textbf{Advantage of a drift-free setup}. Analysis of decorrelation of domain patterns induced by drift for an aperture diameter of (a) \SI{490}{nm} and (b) \SI{2450}{nm}. Data points are correlations of pairs of randomly cropped domain configurations with relative drift distance $d$. The red curve shows the spatial autocorrelation function of the whole domain pattern ($\SI{10}{\micro \meter }\times\SI{10}{\micro \meter}$). Each inset shows an example configuration cropped with the respective aperture size.}
    \label{SI:fig:Simulation_Drift}
\end{figure}

A common issue in long-term XPCS experiments is that instrumental instability, in particular, drift of the sample or illumination may significantly deteriorate the measurements. Drift on the nanometer scale is hard to avoid and leads to a decorrelation without any actual sample dynamics. Heterodyne detection is particularly vulnerable to drift due to its sensitivity to changes of the relative phase between the signal wave and the reference wave. The mask-based geometry proposed here can be realized in an inherently drift-free way when rigorously coupling mask and sample in a monolithic unit. The approach is well known from x-ray imaging via mask-based Fourier-transform holography (FTH) \cite{eisebitt_N_2004, buttner_NP_2015,geilhufe_NC_2014} and was also already used in a quasi-static speckle correlation experiment in Bragg geometry \cite{chen_NC_2019}. Solutions to practically realize the monolithic connection were already developed for FTH and can directly be transferred to mask-based magnetic XPCS. Typically, the actual magnetic samples is produced on the front side of a transparent substrate and the mask is produced into an x-ray-opaque metal film on the back side of the substrate. Methods like focused ion beam milling or electron-beam lithography are used to fabricate the aperture in the mask \cite{buttner_OEO_2013}.

Due to the integrated sample design with a fixed aperture mask, the relative drift between illumination function and the sample is eliminated. If the incoming x-ray beam can be approximated as a plane wave (over the size of the aperture), changes of the overlap between the x-ray beam and the aperture mostly only alter the intensity and absolute phase of the exit wave whereas the relative phase between the sample and the mask apertures is conserved. 

To illustrate how quickly position drift leads to an artificial decorrelation of the signal, we perform Monte-Carlo-like simulations again based on the same maze-domain pattern as before. To this end, we calculate the correlation between two domain configurations selected by a circular aperture defining the illumination function. While the first aperture mask is randomly positioned on the domain pattern, the second mask is slightly displaced by a distance $d$ with respect to the first one to simulate the position drift in between two exposures. Of course, the domain pattern remains static in this simulation. We note that this procedure is conceptually very closely linked to retrieving the spatial autocorrelation of the domain pattern. However, the simulation additionally shows the influence of the size of the probed area restricted by the aperture. 

The simulation is again based on the binary domain pattern from a magnetic force microscope image as used in Sec.~\ref{SI:sec:real_part_approximation}. The magnetic transmission function $\Delta T_\mathrm{magn}$ results from equal absorption and phase contrast ($z_C = 1+\mathrm{i}$) to exclude any errors in calculating the heterodyne correlation function. Binary aperture masks with diameters of  \SI{490}{nm} and \SI{2450}{nm} define the area probed. Drift between signal and reference wave was simulated by shifting the aperture with respect to the magnetic domain pattern. The domain configurations at the initial position and the second, shifted aperture position were correlated using Eq.~\eqref{SI:eq:c_q_2}. For the simulation, 100 random initial positions were chosen and in each case the correlation with 100 randomly shifted positions was evaluated.

The simulated loss of correlation as a function of the position shift $d$ is shown in Fig.~\ref{SI:fig:Simulation_Drift}. On average, the data points from the Monte-Carlo simulation follow the domain's autocorrelation as expected. Neglecting the small (anti-)correlation peaks, the patterns become uncorrelated on average when $d$ approaches the domain width ($\lambda_\mathrm{D}/2$) of the underlying domain structure. As a result, significant decorrelation effects are expected already for drift in the few nanometer regime. More specifically, the correlation drops by \SI{50}{\percent} for a drift distance of only \SI{30}{nm}, assuming a \SI{\sim 100}{nm} texture size as in our case. Such a stability is challenging to achieve in a flexible setup, in particular, for variable sample temperature settings and over a long measurement time, which can take hours in XPCS. Heterodyne XPCS experiments pose even higher challenges as they also rely on long-term stability of the static reference \cite{gutt_PRL_2003,livet_JAC_2007}. In addition, drift also translates to statistical noise as witnessed by the large scatter of the data points, in particular, when the aperture size is small (Fig.~\ref{SI:fig:Simulation_Drift}a).  The mask-based heterodyne approach presented here relaxes the experimental requirements due to the drift-free sample design enabling the detection of nanometer-scale processes. Similar to particle velocity measurements \cite{livet_JAC_2007,ehrburger-dolle_M_2012,ulbrandt_NP_2016}, heterodyning also opens up opportunities to detect correlated translational magnetization dynamics, driven by, e.g., applied magnetic field \cite{pal_S_2018} or spin-polarized electric currents \cite{zazvorka_NN_2019}. 

\section{Conclusions}
We have shown how first-order magnetic correlations can be reconstructed from the heterodyne mixing of charge and magnetic scattering. We suggested a practical implementation that largely reduces errors due to the loss of information of the imaginary part of the scattering signal and that eliminates the influence of drift. Our approach not only enhances the sensitivity and precision of magnetic XPCS measurements, but more generally establishes a direct mathematical link between the Fourier-space correlation signal and the underlying real-space dynamics in the sample. We anticipate broad applications of the technique to study dynamics in magnetic materials.

\section*{Acknowledgments}
Work at MIT was supported by the DARPA TEE program. Work at HZB was supported by the Helmholtz Young Investigator Group Program. This research used resources of the National Synchrotron Light Source II, a US Department of Energy (DOE) Office of Science User Facility operated for the DOE Office of Science by Brookhaven National Laboratory under contract No. DESC0012704.

\bibliography{Theory-Paper}

\end{document}